\documentclass[11pt,onecolumn,twoside]{IEEEtran}
\usepackage{times}
\usepackage{amsbsy}
\usepackage{amsmath}
\usepackage{latexsym}
\usepackage{amssymb}
\usepackage{times}
\usepackage{graphics}
\usepackage[ansinew]{inputenc}
\usepackage[dvips ]{graphicx}
\usepackage[dvips]{pstcol}
\input epsf
\title{Low-Rank Signal Processing: Design, Algorithms for Dimensionality Reduction and Applications
}\vspace{-0.05em}
\author{Rodrigo C. de Lamare    \\ \vspace{-0.05em}
\thanks{Dr. R. C. de Lamare is with the University of York, York Y010 5DD,
UK and with PUC-Rio, Brazil.  E-mails: rcdl500@ohm.york.ac.uk. This work was
supported in part by the UK Ministry of Defence.} }
\date{  }
\begin{document}
\maketitle
\begin{abstract}

We present a tutorial on reduced-rank signal processing, design
methods and algorithms for dimensionality reduction, and cover a
number of important applications. A general framework based on
linear algebra and linear estimation is employed to introduce the
reader to the fundamentals of reduced-rank signal processing and
to describe how dimensionality reduction is performed on an
observed discrete-time signal. A unified treatment of
dimensionality reduction algorithms is presented with the aid of
least squares optimization techniques, in which several techniques
for designing the transformation matrix that performs
dimensionality reduction are reviewed. Among the dimensionality
reduction techniques are those based on the eigen-decomposition of
the observed data vector covariance matrix, Krylov subspace
methods, joint and iterative optimization (JIO) algorithms and JIO
with simplified structures and switching (JIOS) techniques. A
number of applications are then considered using a unified
treatment, which includes wireless communications, sensor and
array signal processing, and speech, audio, image and video
processing. This tutorial concludes with a discussion of future
research directions and emerging topics.

%\begin{keywords}
%%Dimensionality reduction, reduced-rank signal processing
%%techniques, adaptive algorithms, CDMA systems .
%%\end{keywords}
%
\end{abstract}

%\section{Overview:}
%
%We will describe the origins of reduced-rank signal processing,
%provide a literature review, cover the fundamentals, the main
%motivation, goals, and applications.

\section{Introduction}

\PARstart{R}{e}duced-rank signal processing is an area of signal
processing that is strategic for dealing with high-dimensional
data, in low-sample support situations and large optimization
problems that has gained considerable attention in the last decade
\cite{haykin,scharf1}. The origins of reduced-rank signal
processing lie in the problem of feature selection encountered in
statistical signal processing, which refers to a dimensionality
reduction process whereby a data space is transformed into a
feature space \cite{scharf2}. The fundamental idea is to devise a
transformation that performs dimensionality reduction so that the
data vector can be represented by a reduced number of effective
features and yet retain most of the intrinsic information content
of the input data \cite{scharf2}. The goal is to find the best
trade-off between model bias and variance in a cost-effective way,
yielding a reconstruction error as small as desired.

Dimensionality reduction is an emerging and strategic topic that
promises great advances in the fields of statistical signal
processing, linear algebra, communications, multimedia, artificial
intelligence, optimization, control and physics due to its ability
to deal with large systems and to offer high performance at low
computational cost. Central to this idea is the existence of some
form of redundancy in the signals being processed, which allows a
designer to judiciously exploit it by selecting the key features
of the signals. While the data in these applications may be
represented in high dimensions due to the immense capacity for
data retrieval of some systems, the important features are
typically concentrated on lower dimensional subsets—manifolds—of
the measurement space. This allows for significant dimension
reduction with minor or no loss of information. In a number of
applications, the dimensionality reduction may also lead to a
performance improvement due to the denoising property - one
retains the signal subspace and eliminates the noise subspace.
Specifically, this redundancy can be typically characterized by
data that exhibits reduced-rank properties and sparse signals. In
these situations, dimensionality reduction provides a means to
increase the speed and the performance of signal processing tasks,
reduce the requirements for storage and improve the tracking
performance of dynamic signals. This is particularly relevant to
problems which involve large systems, where the design and
applicability of methods is constrained by factors such as
complexity and power consumption.

In general, the dimensionality reduction problem is associated
with reduced-rank operators, characterized by a mapping performed
by an $M \times D$ transformation matrix ${\mathbf S}_D$ with $D <
M$ that compresses the $M \times 1$ observed data vector ${\mathbf
r}$ into a $D \times 1$ reduced-rank data vector ${\mathbf r}_D$.
Mathematically, this relationship is given by ${\mathbf r}_D =
{\mathbf S}_D^H {\mathbf r}$, where $(\cdot )^H$ is the Hermitian
operator. It is desirable to perform these operations so that the
reconstruction error and the computational burden are minimal. The
dimensionality reduction and the system performance are
characterized by (a) accuracy, (b) compression ratio ({\rm CR}),
and (c) complexity. The main challenge is how to efficiently and
optimally design ${\mathbf S}_D$. After dimensionality reduction,
a signal processing algorithm is used to perform the task desired
by the designer. The resulting scheme with $D$ elements can
benefit from a reduced number of parameters, which may lead to
lower complexity, smaller requirements for storage, faster
convergence and better tracking capabilities of time-varying
signals.

In the literature, a number of dimensionality reduction techniques have been
considered based on principal components (PC) analysis
\cite{hotelling}-\cite{ulf}, random projections \cite{kohonen,delamaresp},
diffusion maps \cite{lafon}, incremental manifold learning \cite{law},
clustering techniques \cite{yan,sanguinetti}, Krylov subspace methods that
include the multi-stage Wiener filter (MSWF)
\cite{goldstein,honig,delamare_ccmmswf,song} and the auxiliary vector filtering
(AVF) algorithm \cite{pados,qian}, joint and iterative optimization (JIO)
techniques \cite{hua1,hua,delamare_jio_spl,delamaretvt10,jio_mimo} and JIO
techniques with simplified structures and switching mechanisms (JIOS)
\cite{delamare_jidf}-\cite{barc}. It is well known that the optimal linear
dimensionality reduction transformation is based on the eigenvalue
decomposition (EVD) of the known input data covariance matrix ${\mathbf R}$ and
the selection of the PC. However, this covariance matrix must be estimated. The
approach employed to estimate ${\mathbf R}$ and perform dimensionality
reduction is of central importance and directly affects the performance of the
method. Some methods are plagued by numerical instability, high computational
complexity and large sensitivity to the selected rank $D$. A common and
fundamental limitation of a number of existing methods is that they rely on
estimates of the covariance matrix ${\mathbf R}$ of the data vector ${\mathbf
r}$ to design ${\mathbf S}_D$, which requires a number of data vectors
proportional to the dimension $M$ of ${\mathbf R}$.

The goal of this paper is to provide a tutorial on this important
set of methods and algorithms, to identify important applications
of reduced-rank signal processing techniques as well as new
directions and key areas that deserve further investigation. The
virtues and deficiencies of existing methods will be reviewed in
this article, and a discussion of future trends in the area will
be provided taking into account application requirements such as
the ability to track dynamic signals, complexity and flexibility.
The paper is structured as follows. Section I introduces the
fundamentals of reduced-rank signal processing, the signal model
and the idea of dimensionality reduction using a transformation
matrix. Section II covers the design of the transformation matrix
and a subsequent parameter vector using a least squares approach.
Section III reviews several methods available in the literature
for dimensionality reduction and provides a discussion of their
main advantages and drawbacks. Section IV is devoted to the
applications of these methods, whereas Section V draws the main
conclusions and discusses future research directions.

\section{Fundamentals and signal model}

In this section, our goal is to present the fundamental ideas of
reduced-rank signal processing and how the dimensionality
reduction is performed. We will rely on an approach based on
linear algebra to describe the signal processing of a basic linear
signal model. This model is sufficiently general to account for
numerous applications and topics of interests. Let us consider the
following linear signal model at time instant $i$ that comprises a
set of $M$ samples organized in a vector as given by
\begin{equation}
{\mathbf r}[i] = {\mathbf H} {\mathbf s}[i] + {\mathbf n}[i],~~~
i=1,2, \ldots, P \label{obsig}
\end{equation}
where ${\mathbf r}[i]$ is the $M \times 1$ observed signal vector
which contains the samples to be processed, ${\mathbf H}$ is the
$M \times M$ matrix that describes the mixing nature of the model,
${\mathbf s}[i]$ is the $M \times 1$ signal vector that is
generated by a given source, ${\mathbf n}[i]$ is an $M\times 1$
vector of noise samples, and $P$ is the number of observed signal
vector or simply the data record size.

\begin{figure}[!htb]
\begin{center}
\def\epsfsize#1#2{0.75\columnwidth}
\epsfbox{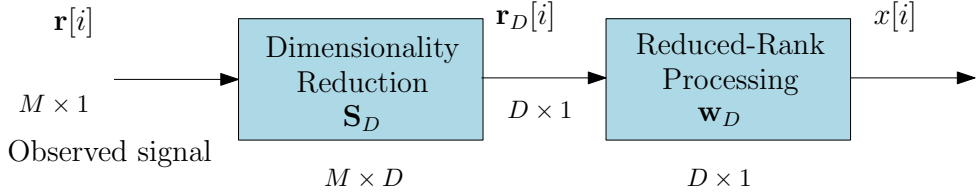} \vspace{0.05em} \caption{Reduced-rank signal
processing stages: dimensionality reduction and reduced-rank
processing.} \label{fig1}
\end{center}
\end{figure}

In reduced-rank signal processing, the main idea is to process the
observed signal ${\mathbf r}[i]$ in two stages, as illustrated in
Fig. 1. The first stage corresponds to the dimensionality
reduction, whereas the second corresponds to the signal processing
in an often low-dimensional subspace. The dimensionality reduction
is performed by a mapping represented by a transformation matrix
${\mathbf S}_D$ with dimensions $M \times D$, where $D < M$, that
projects the observed data vector ${\mathbf r}$ with dimension $M
\times 1$ onto a $D \times 1$ reduced-dimension data vector
${\mathbf r}_D$. This relationship is expressed by
\begin{equation}
{\mathbf r}_D[i] = {\mathbf S}_D^H {\mathbf r}[i] = {\mathbf
S}_D^H\big({\mathbf H} {\mathbf s}[i] + {\mathbf n}[i]\big),
\end{equation}
Key design criteria for the transformation ${\mathbf S}_D$ and the
dimensionality reduction are the reconstruction error, the
computational complexity and the compression ratio ${\rm CR}=M/D$.
These parameters usually depend on the application and the design
requirements.

After the dimensionality reduction, an algorithm is used to
perform the signal processing task on the reduced-dimension
observed vector ${\mathbf r}_D[i]$ according to the designer's
aims. The resulting scheme with $D$ elements will hopefully
benefit from a reduced number of parameters, which may lead to
lower complexity, smaller requirements for storage, faster
convergence and superior tracking capability. In the case of a
combination of weights (filtering) by a parameter vector with $D$
coefficients ${\mathbf w}_D =[ {w}_1 ~{w}_2~\ldots {w}_D]^T$, we
have the following output
\begin{equation}
x[i] = {\mathbf w}^{H}_D{\mathbf r}_D[i]= {\mathbf
w}^{H}_D{\mathbf S}_D^H {\mathbf r}[i].
\end{equation}
It is expected that the output of the reduced-rank signal
processing system will yield a small reconstruction error as
compared to the full-rank system, and provide extra benefits such
as speed of computation and a reduced set of features for
extraction from ${\mathbf r}_D[i]$.

\section{Linear MMSE Design of Reduced-Rank Techniques}

In this section, we will consider a framework for reduced-rank
techniques based on the linear minimum mean-square error (MMSE)
design. The basic idea is to find a reduced-rank model that can
represent the original full-rank model by extracting its key
features. The main goal is to present the design of the main
components employed for reduced-rank processing and examine the
model order selection using a simple and yet general approach. Let
us consider the $M \times 1$ observed signal vector ${\mathbf
r}[i]$ in (\ref{obsig}). For the sake of simplicity and for
general illustrative purposes, we are interested in designing
reduced-rank techniques with the aid of linear MMSE design
techniques. In order to process the data vector ${\mathbf r}[i]$
with reduced-rank techniques, we need to solve the following
optimization problem
\begin{equation}
\begin{split}
\big[{\mathbf S}_{D, {\rm opt}}, {\mathbf w}_{D,{\rm opt}} \big] =
\arg \min_{ {\mathbf S}_{D}, {\mathbf w}_D} E\big[ | d[i] -
\underbrace{{\mathbf w}^{H}_D{\mathbf S}_{D}^H{\mathbf
r}[i]}_{x[i]}|^2 \big], \label{ms1}
\end{split}
\end{equation}
where $d[i]$ is the desired signal and $E\big[\cdot \big]$ stands
for the expected value operator.

The optimal solution ${\mathbf w}_{D,{\rm opt}}$ of the
optimization problem in (\ref{ms1}) is obtained by fixing
${\mathbf S}_D$, taking the gradient terms of the argument with
respect to ${\mathbf w}_D^*$ and equating them to a null vector
\cite{haykin,luen}, which yields
\begin{equation}
\begin{split}
{\mathbf w}_{D,{\rm opt}} & = \bar{\mathbf R}^{-1}\bar{\mathbf p}
= \big({\mathbf S}_D^H{\mathbf R}{\mathbf S}_D\big)^{-1}{\mathbf
S}_D^H{\mathbf p}, \label{opv}
\end{split}
\end{equation}
where $\bar{\mathbf R} = E\big[\bar{\mathbf r}[i]\bar{\mathbf
r}^{H}[i]\big]={\mathbf S}_D^H{\mathbf R}{\mathbf S}_D$ is the $D
\times D$ reduced-rank correlation matrix, $\bar{\mathbf p}=
E\big[ d^*[i]\bar{\mathbf r}[i]\big]={\mathbf S}_D^H{\mathbf p}$
is the $D \times 1$ cross-correlation vector of the reduced-rank
model. The associated MMSE for a rank-$D$ parameter vector is
expressed by
\begin{equation}
\begin{split}
{\rm MMSE} & = \sigma^2_d - \bar{\mathbf p}^H \bar{\mathbf
R}^{-1}\bar{\mathbf p} \\ & = \sigma^2_d - {\mathbf p}^H{\mathbf
S}_D ({\mathbf S}_D^H{\mathbf R}{\mathbf S}_D)^{-1} {\mathbf
S}_D^H{\mathbf p}, \label{mse}
\end{split}
\end{equation}
where $\sigma^2_d=E\big[| d[i]|^2\big]$.

The optimal solution ${\mathbf S}_{D,{\rm opt}}$ of the
optimization problem in (\ref{ms1}) is obtained by fixing
${\mathbf w}_D[i]$, taking the gradient terms of the the
associated MMSE in (\ref{mse}) with respect to ${\mathbf S}_D^*$
and equating them to a zero matrix. By considering the
eigen-decomposition of ${\mathbf R} = {\boldsymbol \Phi}
{\boldsymbol \Lambda} {\boldsymbol \Phi}^H$, where ${\boldsymbol
\Phi}$ is an $M \times M$ unitary matrix with the eigenvectors of
${\mathbf R}$ and ${\boldsymbol \Lambda}$ is an $M \times M$
diagonal matrix with the eigenvalues of ${\mathbf R}$ in
decreasing order, we have
\begin{equation}
{\mathbf S}_{D,{\rm opt}} = {\boldsymbol \Phi}_{1:M,1:D}
\end{equation}
where ${\boldsymbol \Phi}_{1:M,1:D}$ is a $M\times D$ unitary
matrix that corresponds to the signal subspace and contains the
$D$ eigenvectors associated with the $D$ largest eigenvalues of
the unitary matrix ${\boldsymbol \Phi}$. In our notation, the
subscript represents the number of components in each dimension.
For example, the $M\times D$ matrix ${\boldsymbol \Phi}_{1:M,1:D}$
contains the $D$ first columns of ${\boldsymbol \Phi}$, where each
column has $M$ elements.

If we substitute the expression of the optimal transformation $
{\mathbf S}_{D,{\rm opt}} = {\boldsymbol \Phi}_{1:M,1:D}$ and use
the fact that $${\mathbf R} = {\boldsymbol \Phi} {\boldsymbol
\Lambda} {\boldsymbol \Phi}^H =  \big[\underbrace{{\boldsymbol
\Phi}_{1:M,1:D}}_{\rm signal~subspace}~~ \underbrace{{\boldsymbol
\Phi}_{1:M,D+1:M}}_{\rm noise~subspace} \big]
\left[\begin{array}{c c} {\boldsymbol \Lambda}_{1:D,1:D} &
\\ & {\boldsymbol \Lambda}_{D+1:M,D+1:M} \end{array}\right]
\big[{\boldsymbol \Phi}_{1:M,1:D} ~~ {\boldsymbol
\Phi}_{1:M,D+1:M} \big]^H$$ in the expression for the optimal
reduced-rank parameter vector in (\ref{opv}), we have
\begin{equation}
\begin{split}
{\mathbf w}_{D,{\rm opt}} & =  \big({\mathbf S}_D^H{\mathbf
R}{\mathbf S}_D\big)^{-1}{\mathbf S}_D^H{\mathbf p}  =
{\boldsymbol \Lambda}^{-1}_{1:D} {\boldsymbol
\Phi}^H_{1:D}{\mathbf p},
\end{split}
\end{equation}
The development above shows us that the key aspect for
constructing reduced-rank techniques is the design of ${\mathbf
S}_D$ since the MMSE in (\ref{mse}) depends on ${\mathbf p}$,
${\mathbf R}$ and ${\mathbf S}_D$. The quantities ${\mathbf p}$
and ${\mathbf R}$ are common to both reduced-rank and full-rank
designs, however, the matrix ${\mathbf S}_D$ plays a key role in
the dimensionality reduction and in the performance. {The strategy
is to find the most appropriate trade-off between the model bias
and variance \cite{scharf1} by adjusting the rank $D$.

Our exposition assumes so far a reduced-rank linear model with
model order $D$ that is able to represent a full-rank model with
dimension $M$. However, it is well known that the performance,
compression ratio ({\rm CR}) and complexity of such a procedure
depends on the model order $D$. In order to address this problem,
a number of techniques have been reported in the literature which
include the Akaike's information-theoretic (AIC) criterion
\cite{aic}, the minimum description length criterion \cite{mdl}
and a number of other techniques \cite{stoicaspm}. The basic idea
of these methods is to determine the model order $D$ for which a
given criterion is optimized. This can be cast as the following
optimization problem
\begin{equation}
D_{\rm opt} = \arg \min_{D} f(D),
\end{equation}
where $f(D)$ is the objective function that depends on the model
order and consists of a suitable criterion for the problem. This
criterion can be either of an information-theoretic nature
\cite{stoicaspm}, related to the error of the model
\cite{delamare_ccmmswf,delamare_jidf}, associated with metrics or
projections computed from the basis vectors of ${\mathbf S}_D$
\cite{honig} or based on cross-validation techniques \cite{qian}.

\section{Algorithms for Dimensionality Reduction}

For the sake of simplicity and for general illustrative purposes,
we will consider in this section algorithms for dimensionality
reduction based on least squares (LS) optimization techniques with
a fixed model order $D$. In order to process the data vector
${\mathbf r}[i]$ with reduced-rank techniques, we need to solve
the following optimization problem
\begin{equation}
\begin{split}
\big[\hat{\mathbf S}_{D}[i],\hat{\mathbf w}_{D}[i] \big] = \arg
\min_{ {\mathbf S}_{D}[i], {\mathbf w}_D[i]}
\underbrace{\sum_{l=1}^i \lambda^{i-l} | d[l] -
\underbrace{{\mathbf w}^{H}_D[i]{\mathbf S}_{D}^H[i]{\mathbf
r}[l]}_{x[i]}|^2}_{C( {\mathbf S}_D[i], {\mathbf w}_D[i])},
\label{ls1}
\end{split}
\end{equation}
where $d[l]$ is the desired signal and $0< \lambda \leq 1$ stands
for the forgetting factor that is useful for time-varying
scenarios.

The optimal solution $\hat{\mathbf w}_{D}[i]$ of the LS
optimization in (\ref{ls1}) is obtained by fixing ${\mathbf
S}_D[i]$, taking the gradient terms of the argument with respect
to ${\mathbf w}_D^*[i]$ and equating them to a null vector, which
yields
\begin{equation}
\begin{split}
\hat{\mathbf w}_{D}[i] & = {\hat{\bar{\mathbf
R}}}^{-1}[i]{\hat{\bar{\mathbf p}}}[i]\\ & = \big({\mathbf
S}_D^H[i]\hat{\mathbf R}[i]{\mathbf S}_D[i]\big)^{-1}{\mathbf
S}_D^H[i]{\hat{\mathbf p}}[i], \label{woptls}
\end{split}
\end{equation}
where ${\hat{\bar{\mathbf R}}}[i] = \sum_{l=1}^i \lambda^{i-l}
\bar{\mathbf r}[l]\bar{\mathbf r}^{H}[l]={\mathbf
S}_D^H[i]{\hat{\mathbf R}}[i]{\mathbf S}_D[i]$ is the reduced-rank
correlation matrix that is an estimate of the covariance matrix
${\mathbf R}$ and ${\hat{\bar{\mathbf p}}}[i]=\sum_{l=1}^i
\lambda^{i-l} d^*[l]\bar{\mathbf r}[l]={\mathbf
S}_D^H[i]{\hat{\mathbf p}}[i]$ is the cross-correlation vector of
the reduced-rank model that is an estimate of ${\mathbf p}$ at
time $i$. The associated sum of error squares (SES) \cite{haykin}
for a rank-$D$ parameter vector is obtained by substituting
(\ref{woptls}) into the cost function $C( {\mathbf S}_D[i],
{\mathbf w}_D[i])$ in (\ref{ls1}) and is expressed by
\begin{equation}
\begin{split}
{\rm SES} & = \sigma^2_d[i] - {\hat{\bar{\mathbf p}}}^H[i]
{\hat{\bar{\mathbf R}}}^{-1}[i]{\hat{\bar{\mathbf p}}}[i] \\ & =
\sigma^2_d[i] - {\hat{\mathbf p}}^H[i]{\mathbf S}_D[i] ({\mathbf
S}_D^H[i]{\hat{\mathbf R}}[i]{\mathbf S}_D[i])^{-1} {\mathbf
S}_D^H[i]{\hat{\mathbf p}}[i], \label{ses}
\end{split}
\end{equation}
where $\sigma^2_d[i]=\sum_{l=1}^i \lambda^{i-l} | d(l)|^2$. The
development above shows us that the optimal filter $\hat{\mathbf
w}_{D}[i]$ and the SES in (\ref{ses}) depend on $\hat{\mathbf
p}[i]$, $\hat{\mathbf R}[i]$ and ${\mathbf S}_D[i]$. The
quantities $\hat{\mathbf p}[i]$ and $\hat{\mathbf R}[i]$ are
common to both reduced-rank and full-rank designs, however, the
matrix ${\mathbf S}_D[i]$ and the algorithm used for its design
play a key role in dimensionality reduction, the performance and
complexity.

A number of algorithms for computing the transformation ${\mathbf
S}_D[i]$ that is responsible for dimensionality reduction in our
model have been reported in the literature. In this work, we will
categorized them into eigen-decomposition techniques, Krylov-based
methods, joint iterative optimization (JIO) techniques and
techniques based on the JIO with simplified structures and
switching (JIOS). An interesting characteristic that has not been
fully explored in the literature so far is the fact that
algorithms for dimensionality reduction can be devised by looking
at different stages of the signal processing. Specifically, one
can devise algorithms for dimensionality reduction directly from a
cost function associated with the original optimization problem in
(\ref{ls1}) or by considering the SES in (\ref{ses}). In what
follows, we will explore the fundamental ideas behind these
methods and their main features.

\subsection{Eigen-decomposition techniques}

In this part, we review the basic principles of
eigen-decomposition techniques for dimensionality reduction.
Specifically, we focus on algorithms based on the principal
component (PC) and the cross-spectral (CS) approaches. Our aim is
to review the main features, advantages and disadvantages of these
algorithms. The PC was the first statistical signal processing
method employed for dimensionality reduction and was introduced by
Hotelling in the 1930's \cite{hotelling}. The PC method chooses
the subspace spanned by the eigenvectors corresponding to the
largest eigenvalues, which contain the largest portion of the
signal energy. Nevertheless, in the case of a mixture of signals,
the PC approach does not distinguish between the signal of
interest and the interference signal. Hence, the performance of
the algorithm degrades significantly in interference dominated
scenarios. This drawback of the PC algorithm motivated the
development of another technique, the CS algorithm \cite{csmetric}
which selects the eigenvectors such that the MSE over all
eigen-based methods with the same rank is minimal. To do so, it
considers additionally the cross-covariance between the
observation and the unknown signal, thus, being more robust
against strong interference. A common disadvantage of PC and CS
algorithms is the need for eigen-decompositions, which are
computationally very demanding when the dimensions are large and
have typically a cubic cost with $M$ ($O(M)^3$). In order to
address this limitation, numerous subspace tracking algorithms
have been developed in the last two decades, which can reduce the
cost to a quadratic rule with $M$ \cite{Badeau}.

We can illustrate the principle of PC by relying on the framework
employed in this section. The optimal solution ${\mathbf
S}_{D,{\rm opt}}$ of the least squares optimization in (\ref{ls1})
is obtained by fixing ${\mathbf w}_D[i]$, taking the gradient
terms of the the associated SES in (\ref{ses}) with respect to
${\mathbf S}_D^*[i]$ and equating them to a zero matrix. By
considering the eigen-decomposition of $\hat{\mathbf R}[i] =
{\boldsymbol \Phi}[i] {\boldsymbol \Lambda}[i] {\boldsymbol
\Phi}^H[i]$, where ${\boldsymbol \Phi}[i]$ is an $M \times M$
unitary matrix with the eigenvectors of $\hat{\mathbf R}[i]$ and
${\boldsymbol \Lambda}[i]$ is an $M \times M$ diagonal matrix with
the eigenvalues of $\hat{\mathbf R}[i]$ in decreasing order, we
have
\begin{equation}
\hat{\mathbf S}_{D}[i] = {\boldsymbol \Phi}_{1:M,1:D}[i]
\end{equation}
where ${\boldsymbol \Phi}_{1:M,1:D}[i]$ represents the $D$
eigenvectors associated with the $D$ largest eigenvalues of
${\boldsymbol \Phi}[i]$. The adjustment of the model order $D$ of
this method as well as the algorithms described in what follows
can be performed by a model order selection algorithm
\cite{stoicaspm}.

\subsection{Krylov subspace techniques}

The first Krylov methods, namely, the conjugate gradient (CG)
method \cite{cga} and the Lanczos algorithm \cite{lanczos} have
been originally proposed for solving large systems of linear
equations. These algorithms used in numerical linear algebra are
mathematically identical to each other and have been derived for
Hermitian and positive definite system matrices. Other techniques
have been reported for solving these problems and the Arnoldi
algorithm \cite{arnoldi} is a computationally efficient procedure
for arbitrarily invertible system matrices. The multistage Wiener
filter (MSWF) \cite{goldstein} and the auxiliary vector filtering
(AVF) \cite{pados} algorithms are based on a multistage
decomposition of the linear MMSE estimator. A key feature of these
methods is that they do not require an eigen-decomposition and
have a very good performance. It turns out that Krylov subspace
algorithms that are used for solving very large and sparse systems
of linear equations, are suitable alternatives for performing
dimensionality reduction. The basic idea of Krylov subspace
algorithms is to construct the transformation matrix ${\mathbf
S}_D[i]$ with the following structure:
\begin{equation}
\hat{\mathbf S}_{D}[i] = \big[ \hat{\mathbf q}[i] ~ \hat{\mathbf
R}[i]\hat{\mathbf q}[i]~\ldots ~\hat{\mathbf
R}^{D-1}[i]\hat{\mathbf q}[i] \big], \label{kry}
\end{equation}
where $\hat{\mathbf q}[i] = \frac{\hat{\mathbf
p}[i]}{||\hat{\mathbf p}[i]||}$ and $||\cdot||$ denotes the
Euclidean norm (or the $2$-norm) of a vector. In order to compute
the basis vectors of the Krylov subspace (the vectors of
$\hat{\mathbf S}_{D}[i]$), a designer can either directly employ
the expression in (\ref{kry}) or resort to more sophisticated
approaches such as the Arnoldi iteration \cite{arnoldi}. An
appealing feature of the Krylov subspace algorithms is that the
required model order $D$ does not scale with the system size.
Indeed, when $M$ goes to infinity the required $D$ remains a
finite and relatively small value. This result was established by
Xiao and Honig \cite{xiao}. Among the disadvantages of Krylov
subspace methods are the relatively high computational cost of
constructing ${\mathbf S}_{D}[i]$ ($O(DM^2)$), the numerical
instability of some implementations and the lack of flexibility
for imposing constraints on the design of the basis vectors.

\subsection{Joint iterative optimization techniques}

The aim of this part is to introduce the reader to dimensionality
reduction algorithms based on joint iterative optimization (JIO)
techniques. The idea of these methods is to design the main
components of the reduced-rank signal processing system via a
general optimization approach. The basic ideas of JIO techniques
have been reported in \cite{hua1,hua,delamare_jio_spl}. Amongst
the advantages of JIO techniques are the flexibility to choose the
optimisation algorithm and to impose constraints, which provides a
significant advantage over eigen-based and Krylov subspace
methods. One disadvantage that is shared amongst the JIO
techniques, eigen-based and Krylov subspace methods are the
complexity associated with the design of the matrix ${\mathbf
S}_{D}[i]$. For instance, if we are to design a dimensionality
reduction algorithm with a very large $M$, we still have the
problem of having to design an $M \times D$ matrix ${\mathbf
S}_{D}[i]$.

In the framework of JIO techniques, the design of the matrix
${\mathbf S}_{D}[i]$ and the parameter vector ${\mathbf w}_D[i]$
for a fixed model order $D$ will be entirely dictated by the
optimization problem. To this end, we will focus on a generic
${\mathbf S}_{D}[i] = \big[{\mathbf s}_{1}[i]~{\mathbf
s}_{2}[i]~\ldots~ {\mathbf s}_{D}[i]  \big]$, in which the basis
vectors ${\mathbf s}_{d}[i]$, $d=1,2, \ldots, D$ will be obtained
via an optimization algorithm and iterations between the ${\mathbf
s}_{d}[i]$ and ${\mathbf w}_D[i]$ will be performed. The JIO
method consists of solving the following optimization problem
\begin{equation}
\begin{split}
\big[ \hat{\mathbf s}_{1}[i],~ \hat{\mathbf s}_{2}[i],~ \ldots, ~
\hat{\mathbf s}_{D}[i], \hat{\mathbf w}_{D}[i] \big] = \arg \min_{
{\mathbf s}_{1}[i]~{\mathbf s}_{2}[i]~\ldots~ {\mathbf s}_{D}[i],
{\mathbf w}_D[i]} \underbrace{\sum_{l=1}^i \lambda^{i-l} | d[l] -
\underbrace{{\mathbf w}^{H}_D[i] \sum_{d=1}^{D}{\mathbf
s}_{D}^H[i]{\mathbf r}[l]{\mathbf q}_d }_{x[i]}|^2}_{{\mathbf C}
({\mathbf s}_{1}[i],~{\mathbf s}_{2}[i],~\ldots,~ {\mathbf
s}_{D}[i], {\mathbf w}_{D}[i])}, \label{lsjio}
\end{split}
\end{equation}
where ${\mathbf q}_d$ corresponds to a vector with a one in the
$d$th positions and zeros elsewhere. It should be remarked that
the optimization problem in (\ref{lsjio}) is non convex, however,
the algorithms do not present convergence problems. Numerical
studies with JIO methods indicate that the minima are identical
and global. Proofs of global convergence have been established
with different versions of JIO schemes
\cite{hua1,delamare_jio_spl}, which demonstrate that the LS
algorithm converges to the reduced-rank Wiener filter.

\begin{figure}[!htb]
\begin{center}
\def\epsfsize#1#2{0.75\columnwidth}
\epsfbox{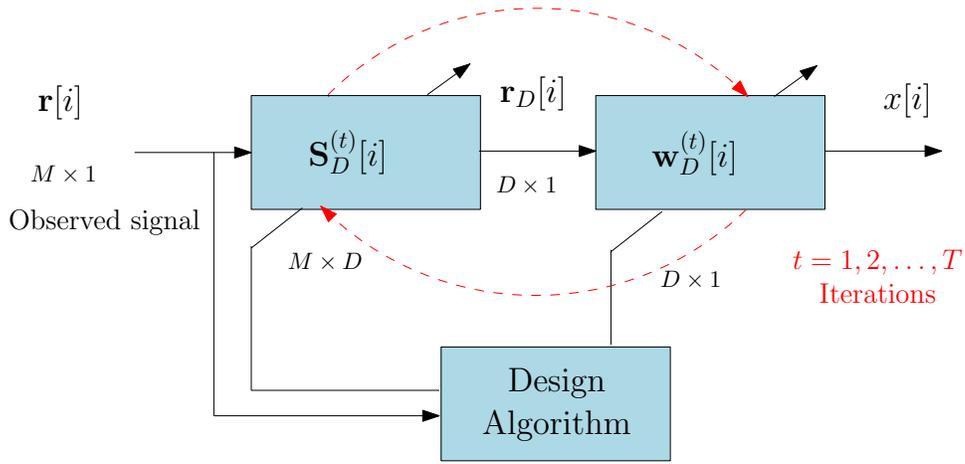} \vspace{0.05em} \caption{JIO scheme:
dimensionality reduction and reduced-rank processing with $T$
iterations.} \label{fig1}
\end{center}
\end{figure}

The solution to the problem in (\ref{lsjio}) for a particular
basis vector $\hat{\mathbf s}_{d}[i]$ for $d=1,2, \ldots, D$ can
be obtained by fixing the remaining basis vectors ${\mathbf
s}_{t}[i]$ for $t \neq d$ and ${\mathbf w}_D[i]$, computing the
gradient terms of the cost function ${\mathbf C} ({\mathbf
s}_{1}[i],~{\mathbf s}_{2}[i],~\ldots,~ {\mathbf s}_{D}[i],
{\mathbf w}_{D}[i])$ defined in (\ref{lsjio}) with respect to
${\mathbf s}_{d}[i]$ and equating the terms to a null vector. The
solution to (\ref{lsjio}) for $\hat{\mathbf w}_D[i]$ can be
obtained in a similar way. We fix all the basis vectors ${\mathbf
s}_{d}[i]$, compute the gradient terms of ${\mathbf C} ({\mathbf
s}_{1}[i],~{\mathbf s}_{2}[i],~\ldots,~ {\mathbf s}_{D}[i],
{\mathbf w}_{D}[i])$ with respect to ${\mathbf w}_{D}[i]$ and
equate the terms to a null vector. Moreover, we also allow the
basis vectors $\hat{\mathbf s}_{d}[i]$ that are the columns of
$\hat{\mathbf S}_D[i]$ and the parameter vector $\hat{\mathbf
w}_{D}[i]$ to exchange information between each other via
$t=1,2,\ldots, T$ iterations, as illustrated in Fig. 2. This
results in the following JIO algorithm
\begin{equation}
\hat{\mathbf s}_{d}^{(t)}[i] =
\frac{1}{|\hat{w}_d^{(t)}[i]|^2}\hat{\mathbf R}^{-1}[i]\big(
\hat{\mathbf p}[i] - \hat{\mathbf v}_d^{(t)}[i] \big),~~~~~ {\rm
for} ~~d=1,2, \ldots, D, ~~ t=1,2, \ldots, T, \label{sdr}
\end{equation}
\begin{equation}
\begin{split}
\hat{\mathbf w}_{D}^{(t)}[i] & =  \big(\hat{\mathbf S}_D^{H,
~(t)}[i]\hat{\mathbf R}[i] \hat{\mathbf
S}_D^{(t)}[i]\big)^{-1}\hat{\mathbf S}_D^{H, ~(t)}[i] \hat{\mathbf
p}[i], ~~~~~ {\rm for}~~ t=1,2, \ldots, T, \label{wdr}
\end{split}
\end{equation}
where $\hat{\mathbf v}_d^{(t)}[i] = \sum_{l=1}^{i} \lambda^{i-l}
{\mathbf r}[l]\hat{w}_d^{(t)}[i] \sum_{n=1, n\neq d}^{D} {\mathbf
r}^H[l] \hat{\mathbf s}_n^{(t)}[i] \hat{w}_n^{(t)}[i]$ is an $M
\times 1$ vector with cross-correlations. The recursions in
(\ref{sdr}) and (\ref{wdr}) are updated over time and iterated $T$
times for each instant $i$ until convergence to a desired
solution. In practice, the iterations can improve the convergence
performance of the algorithms and it suffices to use $T=2,3$
iterations. In terms of complexity, the JIO techniques have a
computational cost that is related to the optimization algorithm.
With recursive LS algorithms the complexity is quadratic with $M$
($(O(M^2)$), whereas the complexity can be as low as linear with
$DM$ when stochastic gradient algorithms are adopted
\cite{delamare_jidf}.

\subsection{Joint iterative optimization techniques with simplified structures and switching}

In this subsection, we introduce the reader to dimensionality
reduction algorithms based on JIO techniques with simplified
structures aided by the concept of switching (JIOS). One
disadvantage that is shared amongst the JIO techniques,
eigen-based and Krylov subspace methods is the complexity
associated with the design of the matrix ${\mathbf S}_{D}[i]$. For
instance, if we are to design a dimensionality reduction algorithm
with a very large $M$, we still have the problem of having to
design an $M \times D$ matrix ${\mathbf S}_{D}[i]$ with the
computational costs associated with $MD$ complex coefficients. An
approach to circumventing this is based on the design of ${\mathbf
S}_{D}[i]$ with simplified structures, which can be done in a
number of ways. For example, a designer can employ random
projections \cite{kohonen} or impose design constraints on
${\mathbf S}_{D}[i]$ such that the number of computations can be
significantly reduced. The main drawback of these simplified
structures is the associated performance loss evidenced by the
large reconstruction error, which is typically larger than that
obtained with more complex dimensionality reduction algorithms. In
order to address this issue, the JIOS framework incorporates
multiple simplified structures that are selected according to a
switching mechanism, aiming at minimizing the reconstruction
error. Switching techniques play a fundamental role in diversity
systems employed in wireless communications systems and control
systems \cite{sun}. They can increase the accuracy of a particular
procedure by allowing a signal processing algorithm to choose
between a number of signals or estimates.

The basic idea of the JIOS-based methods is to address the problem
of reconstruction error associated with the design of a
transformation ${\mathbf S}_{D}[i]$. To this end, the strategy is
to employ multiple transformation matrices in order to obtain
smaller reconstruction errors by seeking the best available
transformation. In order to illustrate the JIOS framework, we
consider the block diagram in Fig. \ref{fig3} in which multiple
transformation matrices in $\hat{\mathbf S}_{D,b}^{(t)}[i]$ for
$b=1,2, \ldots, B$ are employed and iterations can be performed
between $\hat{\mathbf S}_{D,b}[i]$ and the parameter vector
$\hat{\mathbf w}_D^{(t)}[i]$. In addition to this, another goal is
to simplify the structure of $\hat{\mathbf S}_{D,b}[i]$ by
imposing design constraints, which can correspond to having very
few non zero coefficients or even having deterministic patterns
for each branch $b$.

\begin{figure}[!htb]
\begin{center}
\def\epsfsize#1#2{0.75\columnwidth}
\epsfbox{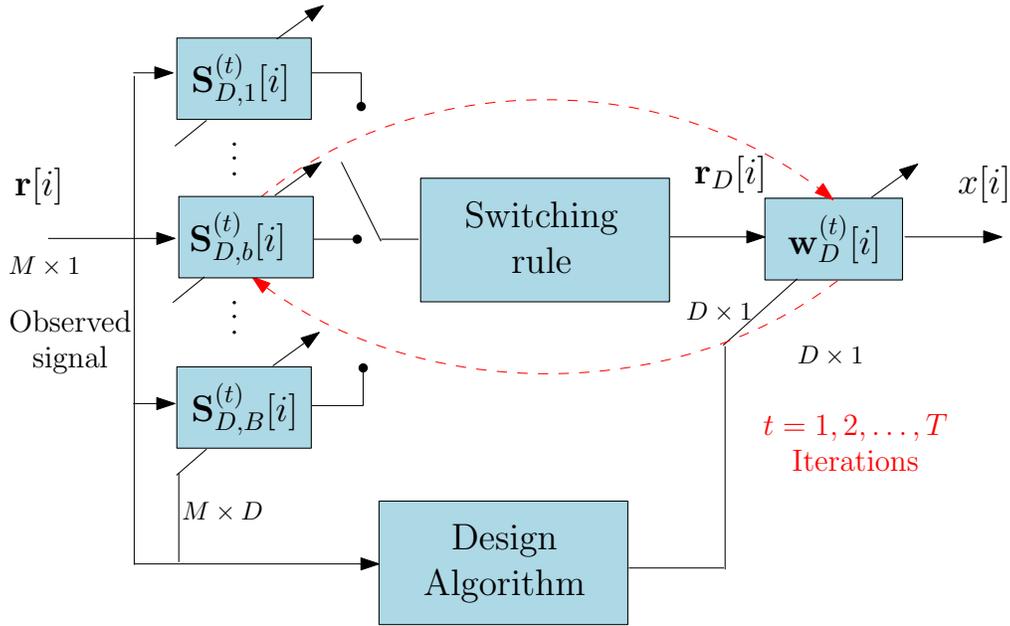} \vspace{0.05em} \caption{JIOS scheme:
dimensionality reduction aided by switching.} \label{fig3}
\end{center}
\end{figure}

One example of the JIOS framework, is the joint and iterative
interpolation, decimation and filtering (JIDF) scheme recently
reported in \cite{delamare_jidf}. Let us now review the JIDF
scheme and describe it using the JIOS framework. In the JIDF
scheme, the basic idea is to employ an interpolator ${\boldsymbol
v}[i]$ with $I$ coefficients, a decimation unit and a reduced-rank
parameter vector ${\boldsymbol w}_D[i]$ with $D$ coefficients. A
key strategy for the design of parameters of the JIDF scheme is to
express the output $x[i] = \bar{\mathbf w}^{H}_D[i]{\mathbf
S}_{D,b}^H[i] {\mathbf r}[i]$  as a function of ${\mathbf v}[i]$,
the decimation matrix ${\mathbf D}_{b}[i]$ and $\bar{\mathbf
w}_D[i]$ as follows:
\begin{equation}
\begin{split}
x[i] & = \bar{\mathbf w}^{H}_D[i] {\mathbf S}_{D,b}^H[i]{\mathbf
r}[i] = \bar{\mathbf w}^{H}_D[i] {\mathbf
D}_{b}[i]\boldsymbol{\Re}_{o}[i] {\mathbf v}^{*}[i] \\ & =
\bar{\mathbf w}^{H}_D[i] \big(\underbrace{{\mathbf D}_{b}[i]
\sum_{m=1}^{M} {\mathbf B}_m {\mathbf v}^*[i] }_{{\mathbf
S}_{D,b}^H[i]} \big) {\mathbf r}[i] = {\mathbf v}^{H}[i]{\mathbf
u}[i]
\end{split}, \label{output}
\end{equation}
where ${\mathbf u}[i]= \boldsymbol{\Re}_{o}^T[i]{\mathbf
D}^T_{b}[i] \bar{\mathbf w}^{*}[i]$ is an ${I}\times 1$ vector,
the $M \times I$ matrix ${\boldsymbol B}_m$ has an $I$-dimensional
identity matrix starting at the $m$-th row, is shifted down by one
position for each $m$ and the remaining elements are zeros, and
the $M\times I$ matrix ${\boldsymbol \Re}_{o}[i]$ with the samples
of ${\mathbf r}[i]$ has a Hankel structure described by
\begin{equation}
{\boldsymbol \Re}_{o}[i] = \left[\begin{array}{c c c c c}
r_{0}^{[i]} & r_{1}^{[i]}   & \ldots & r_{{I}-1}^{[i]}  \\
r_{1}^{[i]}  & r_{2}^{[i]}   & \ldots & r_{{I}}^{[i]}  \\
\vdots & \vdots  & \ddots & \vdots \\
r_{M-1}^{[i]}  & r_{M}^{[i]}  & \ldots & r_{M +{I}-2}^{[i]}  \\
 \end{array}\right].
\end{equation}
The expression in (\ref{output}) indicates that the dimensionality
reduction carried out by the proposed scheme depends on finding
appropriate ${\mathbf v}[i]$, ${\mathbf D}_{b}[i]$ for
constructing ${\mathbf S}_{D,b}[i]$.

The design of the decimation matrix ${\mathbf D}_b[i]$ employs for
each row the structure:
\begin{equation}
{\mathbf d}_{j,b}[i] = [\underbrace{0~~ \ldots ~~
0}_{\gamma_{j}~zeros} ~~~ 1 ~~  \underbrace{ 0  ~~~ \ldots ~~
0}_{(M-\gamma_{j}-1) ~zeros} ]^T, \label{decvec}
\end{equation}
and the index $j$ ($j=1,2,\ldots, D$) denotes the $j$-th row of
the matrix, the rank of the matrix ${\mathbf D}_b[i]$ is $D=M/L$,
the decimation factor is $L$ and $B$ corresponds to the number of
parallel branches. The quantity $\gamma_{j}$ is the number of
zeros chosen according to a given design criterion.Given the
constrained structure of ${\boldsymbol D}_b[i]$, it is possible to
devise an optimal procedure for designing ${\boldsymbol D}_b[i]$
via an exhaustive search of all possible design patterns with the
adjustment of the variable $\gamma_{j}$. The exhaustive procedure
has a total number of patterns equal to ${ B}_{\rm ex} =  \binom
{M} {D}$. The exhaustive scheme is too complex for practical use
and it is fundamental to devise decimation schemes that are
cost-effective. By adjusting the variable $\gamma_{j}$, a designer
can obtain various sub-optimal schemes including pre-stored (e.g.
$\gamma_{j} = (j-1)L+(b-1)$) and random patterns.

The decimation matrix ${\mathbf D}_b[i]$ is selected to minimize
the square of the instantaneous error signal obtained for the $B$
branches employed as follows
\begin{equation}
{\mathbf D}_b[i] = {\mathbf D}_{b_{\rm s}}[i] ~~ \textrm{when} ~~
b_{\rm s} = \arg \min_{1\leq b \leq B} |e_{b}[i]|^{2},
\label{Ddesign2}
\end{equation}
where $ e_{b}[i] = d[i]- \bar{\mathbf w}^{H}_D[i]{\mathbf
S}_{D,b}^H[i]{\mathbf r}[i]$. The $D \times 1$ vector ${\mathbf
r}_D[i]$ is computed by
\begin{equation}
{\mathbf r}_{D}[i] = {\mathbf S}_{D,b}^H[i]{\mathbf r}[i]
={\mathbf D}_{b}[i] {\boldsymbol \Re}_{o}[i]{\mathbf v}^{*}[i],
\end{equation}
The design of ${\mathbf S}_{D,b}[i]$ and ${\mathbf w}_D[i]$
corresponds to solving the following optimization problem
\begin{equation}
\begin{split}
\big[\hat{\mathbf S}_{D}[i], \hat{\mathbf w}_{D}[i] \big] = \arg
\min_{ {\mathbf v}[i],  {\mathbf w}_D[i]} \sum_{l=1}^i
\lambda^{i-l} | d[l] - \bar{\mathbf w}^{H}_D[i] {\mathbf
D}_{b}[l]\boldsymbol{\Re}_{o}[l] {\mathbf v}^{*}[i]|^2,
\label{lsjidf}
\end{split}
\end{equation}
It should be remarked that the optimization problem in
(\ref{lsjidf}) is non convex, however, the methods do not present
problems with local minima. Proofs of convergence are difficult
due to the switching mechanism and constitute an interesting open
problem. Using a similar approach to the JIO technique, we obtain
the following recursions for computing ${\mathbf v}[i]$ and
${\mathbf w}_D[i]$:
\begin{equation}
\hat{\mathbf v}^{(t)}[i] = \hat{\mathbf R}^{-1}_{ u}[i]
\hat{\mathbf p}_{ u}^{(t)}[i], ~~ t=1,2,\ldots, T \label{vrec}
\end{equation}
\begin{equation}
\hat{\mathbf w}^{(t)}_D[i] = \hat{\mathbf R}^{-1}_{D}[i]
\hat{\mathbf p}_{D}^{(t)}[i], ~~ t=1,2,\ldots, T \label{vrec}
\end{equation}
where $\hat{\mathbf p}_{ u}^{(t)}[i] = \sum_{l=1}^{i}
\lambda^{i-l} d^*[l]{\mathbf u}^{(t)}[l]$, $\hat{\mathbf R}_{
u}[i]= \sum_{l=1}^{i} \lambda^{i-l} {\mathbf u}^{(t)}[l]{\mathbf
u}^{H,~(t)}[l]$, $\hat{\mathbf p}_{D}^{(t)}[i]= \sum_{l=1}^{i}
\lambda^{i-l} d^*[l]{\mathbf r}_D^{(t)}[l]$ and $\hat{\mathbf
R}_{D}^{(t)}[i]= \sum_{l=1}^{i} \lambda^{i-l} {\mathbf
r}_D^{(t)}[l]{\mathbf r}^{H^,~(t)}_D[l]$. In terms of complexity,
the computational cost of the JIDF scheme scales linearly with $D
I M$ and quadratically with $I$ and $D$. Since $I$ and $D$ are
typically very small ($3$-$5$ coefficients), this makes the JIDF
scheme a low-complexity alternative as compared with PC, Krylov
and JIO schemes.

\subsection{Summary of Dimensionality Reduction Algorithms}

In this part, we summarize the most representative LS-based
algorithms for dimensionality reduction presented in the previous
subsections and provide a table that explain how to simulate these
algorithms. Specifically, we consider the PC method
\cite{jolliffe}, the MSWF \cite{honig}, the JIO
\cite{delamare_jio_spl} and the JIDF techniques
\cite{delamare_jidf}.

{\footnotesize \linespread{1}
\begin{table}[htb!]
\flushleft %centering
 \caption{\normalsize Dimensionality reduction algorithms.} {
\begin{tabular}{ll}
PC : \rule{0pt}{2.6ex} \rule[-1.2ex]{0pt}{0pt}&\\
\hline Step 1: & Initialization:\rule{0pt}{2.6ex} \\
& Set values for $\lambda$, the model order $D$ and $\hat{\mathbf R}[0]$\\

Step 2: & For i= 1, 2, \ldots, P.\\
&
\begin{tabular}{ll}
 (1) Compute $\hat{\mathbf R}[i] = \lambda \hat{\mathbf R}[i-1] + {\mathbf r}[i]{\mathbf r}^H[i]$,\\
 (2) Compute $\hat{\mathbf p}[i] = \lambda \hat{\mathbf p}[i-1] + {\mathbf p}[i]{ d}^*[i]$,\\
 (3) Perform an eigen-decomposition of $\hat{\mathbf R}[i]= {\mathbf \Phi}[i] {\mathbf \Lambda}[i]{\mathbf \Phi}^H[i]$ , \\
 (4) Compute $\hat{\mathbf S}_D[i] = {\mathbf \Phi}_{1:M,1:D}[i]$, \\
 (5) Calculate $\hat{\mathbf w}_D[i] = \big(\hat{\mathbf S}_D^H[i]\hat{\mathbf R}[i]\hat{\mathbf S}_D[i]\big)^{-1}\hat{\mathbf S}_D^H[i]\hat{\mathbf p}[i]$.\\
 %\hline
\end{tabular}
\end{tabular}
\begin{tabular}{ll}
MSWF : \rule{0pt}{2.6ex} \rule[-1.2ex]{0pt}{0pt}&\\
\hline Step 1: & Initialization:\rule{0pt}{2.6ex} \\
& Set values for $\lambda$, the model order $D$ and $\hat{\mathbf R}[0]$\\

Step 2: & For i= 1, 2, \ldots, P.\\
&
\begin{tabular}{ll}
 (1) Compute $\hat{\mathbf R}[i] = \lambda \hat{\mathbf R}[i-1] + {\mathbf r}[i]{\mathbf r}^H[i]$,\\
 (2) Compute $\hat{\mathbf p}[i] = \lambda \hat{\mathbf p}[i-1] + {\mathbf r}[i]{ d}^*[i]$,\\
 (3) Construct $\hat{\mathbf S}_{D}[i] = \big[ \hat{\mathbf q}[i] ~\hat{\mathbf R}[i]\hat{\mathbf
  q}[i]~\ldots ~\hat{\mathbf R}^{D-1}[i]\hat{\mathbf q}[i] \big]$, \\ where $\hat{\mathbf q}[i] = \frac{\hat{\mathbf p}[i]}{||\hat{\mathbf p}[i]||}$,\\
 (4) Calculate $\hat{\mathbf w}_D[i] = \big(\hat{\mathbf S}_D^H[i]\hat{\mathbf R}[i]\hat{\mathbf S}_D[i]\big)^{-1}\hat{\mathbf S}_D^H[i]\hat{\mathbf p}[i]$.\\
 %\hline
\end{tabular}
\end{tabular}
\begin{tabular}{ll}
JIO : \rule{0pt}{2.6ex} \rule[-1.2ex]{0pt}{0pt}&\\
\hline Step 1: & Initialization:\rule{0pt}{2.6ex} \\
& Set values for $\lambda$, the model order $D$ and ${\mathbf R}[0]$\\

Step 2: & For i= 1, 2, \ldots, P.\\
&
\begin{tabular}{ll}
 (1) Compute $\hat{\mathbf R}[i] = \lambda \hat{\mathbf R}[i-1] + {\mathbf r}[i]{\mathbf r}^H[i]$,\\
 (2) Obtain $\hat{\mathbf p}[i] = \lambda \hat{\mathbf p}[i-1] + {\mathbf r}[i]{ d}^*[i]$,\\
 (3) For ~$t=1,2, \ldots, T$ and ~$d=1,2, \ldots, D$\\
 (4) Compute $\hat{\mathbf s}_{d}^{(t)}[i] = \frac{1}{|\hat{w}_d^{(t)}[i]|^2}\hat{\mathbf R}^{-1}[i] \big( \hat{\mathbf p}[i] - \hat{\mathbf v}_d^{(t)}[i] \big)$,  \\
 (5) Compute $\hat{\mathbf w}_{D}^{(t)}[i] =  \big(\hat{\mathbf S}_D^{H,~(t)}[i]\hat{\mathbf R}[i]\hat{\mathbf S}_D^{(t)}[i]\big)^{-1}\hat{\mathbf S}_D^{H, ~(t)}[i]\hat{\mathbf p}[i]$,\\
  where $\hat{\mathbf v}_d^{(t)}[i] = \sum_{l=1}^{i} \lambda^{i-l}{\mathbf r}[l]\hat{w}_d^{(t)}[i] \sum_{n=1, n\neq d}^{D} {\mathbf r}^H[l] \hat{\mathbf s}_n^{(t)}[i] \hat{w}_n^{(t)}[i]$. \\
 %\hline
\end{tabular}
\end{tabular}
\begin{tabular}{ll}
JIO : \rule{0pt}{2.6ex} \rule[-1.2ex]{0pt}{0pt}&\\
\hline Step 1: & Initialization:\rule{0pt}{2.6ex} \\
& Set values for $\lambda$, the model orders $D$ and $I$, $\hat{\mathbf R}_u[0]$ and $\hat{\mathbf R}_D[0]$\\

Step 2: & For $i= 1, 2, \ldots, P$ and $t=1,2, \ldots T$.\\
&
\begin{tabular}{ll}
 (1) Determine ${\mathbf D}_b[i] = {\mathbf D}_{b_{\rm s}}[i] ~~ \textrm{when} ~b_{\rm s} = \arg \min_{1\leq b \leq B} |e_{b}[i]|^{2}$, \\
 (2) Calculate ${\mathbf u}^{(t)}[i]= \boldsymbol{\Re}_{o}^T[i]{\mathbf D}^T_{b}[i] {\hat{\bar{\mathbf w}}}^{*,~(t)}_D[i]$ and ${\mathbf r}_D^{(t)}[i] = \hat{\mathbf S}_D^{(t), ~H}[i] {\mathbf r}[i]$,  \\
 (3) Compute $\hat{\mathbf R}_{ u}[i] = \lambda \hat{\mathbf R}_{ u}[i-1] + {\mathbf u}^{(t)}[i]{\mathbf u}^{(t), ~H}[i]$ and $\hat{\mathbf p}_{ u}^{(t)}[i] = \lambda \hat{\mathbf p}_{ u}^{(t)}[i-1] +  d^*[i]{\mathbf u}^{(t)}[i]$,\\
 (4) Compute $\hat{\mathbf R}_{ D}[i] = \lambda \hat{\mathbf R}_{ D}[i-1] + {\mathbf r}^{(t)}_D[i]{\mathbf r}^{(t),~H}_D[i]$ and $\hat{\mathbf p}_{ D}^{(t)}[i] = \lambda \hat{\mathbf p}_{ D}^{(t)}[i-1] +  d^*[i]{\mathbf r}^{(t)}_{D}[i]$,\\
 (5) Obtain $\hat{\mathbf v}^{(t)}[i] = \hat{\mathbf R}^{-1}_{ u}[i] \hat{\mathbf p}_{u}^{(t)}[i]$,  \\
 (6) Construct $\hat{\mathbf S}_{D}^{(t), ~H}[i] {\mathbf D}[i] \sum_{m=1}^{M}{\mathbf B}_m \hat{\mathbf v}^{*, ~(t)}[i]$, \\
 (7) Obtain $\hat{\mathbf w}^{(t)}_D[i] = \hat{\mathbf R}^{-1}_{D}[i] \hat{\mathbf p}_{D}^{(t)}[i]$,\\
 %\hline
\end{tabular}
\end{tabular}
 } \label{tab:summary}
\end{table}
}

\section{Applications:}

In this section, we will describe a number of applications for
reduced-rank signal processing and dimensionality reduction
algorithms and link them with new research directions and emerging
fields. Among the key areas for these methods are wireless
communications, sensor and array signal processing, speech and
audio processing, image and video processing. A key aspect that
will considered is the need for dimensionality reduction and
typical values for the dimension $M$ of the observed vector, the
model order $D$ and the compression ratio ${\rm CR}=M/D$.

\subsection{ Wireless communications:}

In wireless communications, a designer must deal with stringent
requirements in terms of quality of service, an increasingly
demand for higher data rates and scenarios with time-varying
channels. At the heart of the problems in wireless communications
lie the need for designing transmitters and receivers, algorithms
for data detection and channel and parameter estimation. These
problems are ubiquitous and common to spread spectrum,
multi-carrier and multiple antenna systems. Specifically, when the
number of parameters grows beyond a certain level (which is often
the case), the level of interference is high and the channel is
time-varying, reduced-rank signal processing and algorithms for
dimensionality reduction can play a decisive role in the design of
wireless communication systems. For instance, in spread spectrum
systems we often encounter receiver design problems that require
the computation of a few hundred parameters, i.e., $M=100,200,
\ldots$ and we typically wish to perform a dimensionality
reduction which leads to a model order of a few elements, i.e.,
$D=4,5$ and which yields a ${\rm CR}>25$. In a multi-antenna
system in the presence of flat fading, the number of parameters is
typically small ($2,4$) for spatial processing and corresponds to
the number of antennas used at the transmitter and at the
receiver. However, if we consider multi-antenna systems in the
presence of frequency selective channels, the numbers can increase
to dozens of coefficients. In addition, the combination of
multi-antenna systems with multi-carrier transmissions (eg.
MIMO-OFDM) in the presence of time-selective channels, can require
the equalization of structures with hundreds of coefficients.
Therefore, the use of reduced-rank signal processing techniques
and dimensionality reduction algorithms can be of fundamental
importance in the problems previously described. In what follows,
we will illustrate with a numerical example the design of a
space-time reduced-rank linear receiver for interference
suppression in a spread spectrum system equipped with antenna
arrays.

{\bf \it Numerical Example}: Space-Time Interference Suppression
for Spread Spectrum Systems

We consider the uplink of a direct-sequence code-division multiple
access (DS-CDMA) system with symbol interval $T$, chip period
$T_c$, spreading gain $N=T/Tc$, $K$ users, multipath channels with
the maximum number of propagation paths $L$, where $L < N$. The
system is equipped with an antenna that consists of a uniform
linear array (ULA) and $J$ sensor elements.
%In the model adopted, the intersymbol
%interference (ISI) span and contribution are functions of the
%processing gain $N$ and $L$. For instance, we assume that $L <
%N$ which results in the interference between $3$ symbols in total,
%the current one, the previous and the successive symbols.
The spacing between the ULA elements is $d=\lambda_c/2$, where
$\lambda_c$ is carrier wavelength. We assume that the channel is
constant during each symbol, the base station receiver is
perfectly synchronized and the delays of the propagation paths are
multiples of the chip rate. The received signal after filtering by
a chip-pulse matched filter and sampled at the chip period yields
the $M\times 1$ received vector at time $i$
\begin{equation}
\begin{split}
{\mathbf r}[i] & =  {\mathbf H}[i] {\mathbf s}[i] + {\mathbf n}[i]
\\ & = \sum_{k=1}^{K} A_{k}s_{k}[i-1] \bar{\mathbf p}_{k}[i] +
A_{k}s_{k}[i] {\mathbf p}_{k}[i]  + A_{k}s_{k}[i+1] \tilde{\mathbf
p}_{k}[i] + {\boldsymbol \eta}[i] + {\mathbf
 n}[i],
\end{split}
\end{equation}
where $M=J(N+L-1)$, $s_k[i]$ denotes the data symbol of user $k$
at time $i$, $A_k$ is the amplitude of user $k$, the complex
Gaussian noise vector is ${\mathbf n}[i] = [n_{1}[i]
~\ldots~n_{M}[i]]^{T}$ with $E[{\mathbf n}[i]{\mathbf n}^{H}[i]] =
\sigma^{2}{\mathbf I}$, $(\cdot)^{T}$, ${\boldsymbol \eta}[i]$
corresponds to the intersymbol interference and $(\cdot)^{H}$
denote transpose and Hermitian transpose, respectively, and
$E[\cdot]$ stands for expected value. The spatial signatures for
previous, current and future data symbols are $\bar{\mathbf
p}_{k}[i-1]$, ${\mathbf p}_{k}[i]$, $\tilde{\mathbf p}_{k}[i+1]$
and are constructed with the stacking of the convolution between
the signature sequence ${\mathbf s}_{k} = [a_{k}(1) \ldots
a_{k}(N)]^{T}$ of user $k$ and the $L\times 1$ channel vector
${\mathbf h}_{k,j}[i] = [h_{k,0}^{(l)}[i] \ldots
h_{k,L-1}^{(l)}[i]]^{T}$ of user $k$ at each
antenna element $j=1,2,\ldots,J$. %The spatial signatures for

A linear reduced-rank space-time receiver for a desired user $k$
can be designed by linearly combining the received vector
${\mathbf r}[i]$ with the transformation ${\mathbf S}_D[i]$ and
the reduced-rank parameter vector ${\mathbf w}_D[i]$ as expressed
by
\begin{equation}
x_k[i] = {\mathbf w}_D^H[i]{\mathbf S}_D^H[i]{\mathbf r}[i],
\end{equation}
where the data detection corresponds to applying the signal
$x_k[i]$ to a decision device as given by $\hat{s}_k[i] = Q (
x_k[i])$, where $Q(\cdot)$ is the function that implements the
decision device and depends on the modulation.

Let us now consider simulation examples of the space-time
reduced-rank receiver described above in which the dimensionality
reduction techniques using LS optimization are compared. In all
simulations, we use the initial values ${\bf w}_D[0]=
[1~0~\ldots~0]^T$ and ${\bf S}_D[0]=[{\bf I}_D ~ {\bf
0}_{D,M-D}]^T$, $\lambda=0.998$, employ randomly generated
spreading codes with $N=16$, assume $L=9$ as an upper bound, use
$3$-path channels with relative powers given by $0$, $-3$ and $-6$
dB, where in each run the spacing between paths is obtained from a
discrete uniform random variable between $1$ and $2$ chips and
average the experiments over $1000$ runs. The system has a power
distribution among the users for each run that follows a
log-normal distribution with associated standard deviation equal
to $1.5$ dB. We consider antenna arrays equipped with $J=2$ and
$J=4$ elements. The dimension $M$ of the observed signal in these
examples corresponds to $M=J(N+L-1)$, which corresponds to $M=48$
(for $J=2$) and $M=96$ (for $J=4$). These figures leads to a $CR
>20$ (for $J=4$).  We compare the full-rank, the PC method
\cite{jolliffe}, the MSWF \cite{goldstein} Krylov subspace
technique, the JIO technique \cite{delamare_jio_spl} and the JIOS
(JIDF) scheme with an optimized model order $D$ for each method.
We also include the linear full-rank MMSE receiver that assumes
the knowledge of the channels and the noise variance at the
receiver. Each reduced-rank scheme provides an estimate of the
desired symbol for the desired used (user $1$ in all experiments)
and we assess the bit error rate (BER) against the number of
symbols. We transmit packet with $P=1500$ QPSK symbols in which
$250$ are used for training. After the training phase the
space-time receivers are switched to decision-directed mode. In
Fig. \ref{berxsym}, we show an example of the BER performance
versus the number of symbols, whereas in Fig. \ref{berxsnr&k} we
consider examples with the BER performance versus the SNR and the
number of users.

\begin{figure}[!htb]
\begin{center}
\def\epsfsize#1#2{0.75\columnwidth}
\epsfbox{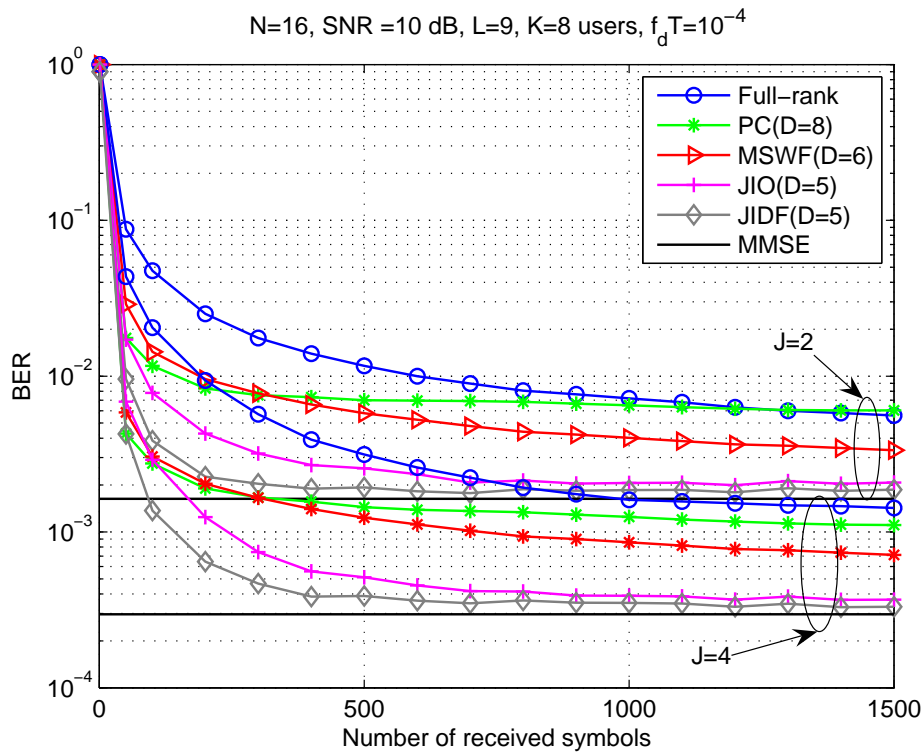} \caption{\small BER performance versus number of received
symbols.} \label{berxsym}
\end{center}
\end{figure}

The curves depicted in Figs. \ref{berxsym} and \ref{berxsnr&k}
show that the dimensionality reduction applied to the received
signals combined with the reduced-rank processing can accelerate
significantly the convergence of the adaptive receivers. The best
results are obtained by the JIDF and JIO methods, which approach
the linear MMSE receiver and are followed by the MSWF, the
PC-based receiver and the full-rank technique. The algorithms
analyzed show very good performance for different values of SNR
and number of users in the system. A key feature to be remarked is
the ability of the subspace-based algorithms to converge faster
and to obtain good results in short data records, reducing the
requirements for training.

\begin{figure}[!htb]
\begin{center}
\def\epsfsize#1#2{0.75\columnwidth}
\epsfbox{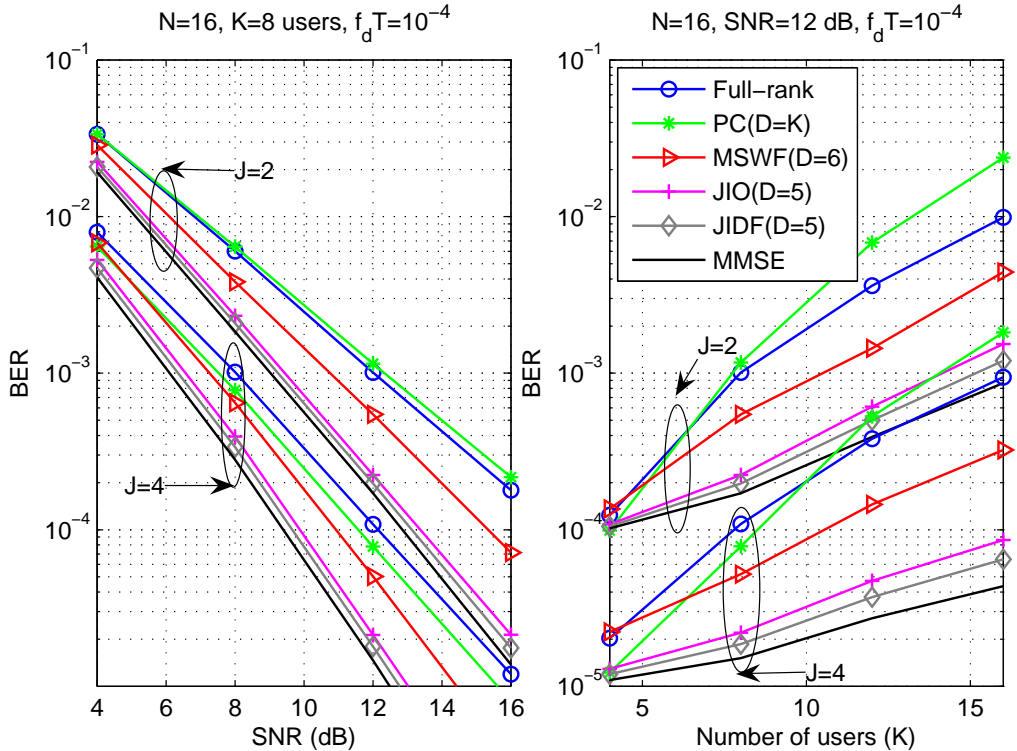} \caption{BER performance against (a) $E_b/N_0$ (dB) and (b)
Number of Users (K) .} \label{berxsnr&k}
\end{center}
\end{figure}

\subsection{Sensor and array signal processing:}

The basic aim of sensor and array signal processing is to consider
temporal and spatial information, captured by sampling a wave
field with a set of appropriately placed antenna elements or
sensor devices. These devices are organized in patterns or arrays
which are used to detect signals and to determine information
about them. The wavefield is assumed to be generated by a finite
number of emitters, and contains information about signal
parameters characterizing the emitters. A number of applications
for sensor and array signal processing have emerged in the last
decades and include active noise and vibration control,
beamforming, direction finding, harmonic retrieval, distributed
processing for networks, radar and sonar systems. In these
applications, when the number of parameters grows beyond a certain
level the signal-to-noise ratio is low and the level of
interference is high, reduced-rank signal processing and
algorithms for dimensionality reduction can offer an improved
performance as compared with conventional full-rank techniques.
For example, in broadband beamforming or space-time adaptive
processing applications for radar we may have to deal with an
optimization problem that requires the processing of observed
signals with dozens to hundreds of coefficients, i.e., $50<M<200$.
In this case, a dimensionality reduction which leads to a model
order with only a few elements, i.e., $D=4,5$ and which has a
${\rm CR}>10$ can facilitate the design and be highly beneficial
to the performance of the system. A number of other problems in
sensor and array signal processing can be cost-effectively
addressed with reduced-rank signal processing techniques and
dimensionality reduction algorithms. In what follows, we will
illustrate with a numerical example the design of adaptive
beamforming techniques.

{\bf \it Numerical Example}: Adaptive reduced-rank beamforming

Let us consider a smart antenna system equipped with a uniform
linear array (ULA) of $M$ elements. Assuming that the sources are
in the far field of the array, the signals of $K$ narrowband
sources impinge on the array $\left( K < M \right)$ with unknown
directions of arrival (DOA) ${\theta}_l$ for $l=1,2, \ldots, K$.
The input data from the antenna array can be organized in an $M
\times 1$ observed vector expressed by
\begin{equation}
\begin{split}
{\mathbf r}[i] & = {\mathbf H}{\mathbf s}[i] + {\mathbf n}[i] \\
& = {\mathbf A}(\theta) {\mathbf s}[i] + {\mathbf n}[i]
\end{split}
\end{equation}
where ${\mathbf A}(\theta) = \left[ {\mathbf a}(\theta_1), \ldots,
{\mathbf a}(\theta_K)\right]$ is the $M \times K$ matrix of signal
steering vectors.  The $M \times 1$ signal steering vector is
defined as
\begin{equation}
{\mathbf a}(\theta_l) = \left[ 1, e^{-2\pi j
\frac{d_s}{\lambda_c}\cos \theta_l}, \ldots, e^{-2\pi j
(M-1)\frac{d_s}{\lambda_c}\cos \theta_l}\right] ^T
\end{equation}
for a signal impinging at angle $\theta_l$, $l=1,2, \ldots, K$,
where $d_s = \lambda_c / 2$ is the inter-element spacing,
$\lambda_c$ is the wavelength and $(.)^T$ denotes the transpose
operation. The vector ${\mathbf n}[i]$ denotes the complex vector
of sensor noise, which is assumed to be zero-mean and Gaussian
with covariance matrix $\sigma^2 {\mathbf I}$.

Let us now consider the design of an adaptive reduced-rank minimum
variance distortionless response (MVDR) beamformer. The design
problem consists of computing the transformation ${\mathbf
S}_D[i]$ and ${\mathbf w}[i]$ via the following optimization
problem
\begin{equation}
\begin{split}
\big[ \hat{\mathbf S}_{D}[i], \hat{\mathbf w}_{D}[i] \big] & =
\arg \min_{{\mathbf S}_D[i], {\mathbf w}_D[i]} ~ {\mathbf
w}_D^{H}[i] {\mathbf S}_D^H[i] {\boldsymbol R}[i] {\mathbf
S}_D[i] {\mathbf w}_D[i] \\
{\textrm {subject to}}  ~ {\mathbf w}^H_D[i]{\mathbf S}_D^H[i]
{\boldsymbol a}(\theta_k) & = 1 \label{propt}
\end{split}
\end{equation}
In order to solve the above problem, we can resort to algorithms
outlined in Section IV. The main difference is that now we need to
minimize the mean-square value of the output of the array and
enforce the constraint that ensures the response of the
reduced-rank beamforming algorithm to be equal to unity.

\begin{figure}[!htb]
\begin{center}
\def\epsfsize#1#2{0.75\columnwidth}
\epsfbox{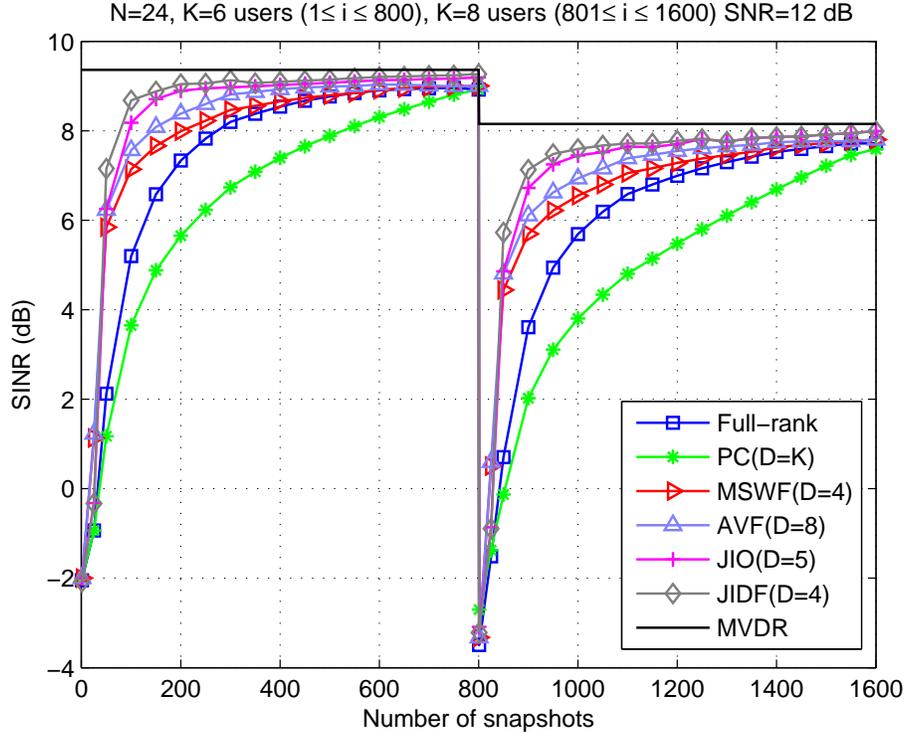} \caption{\small SINR performance versus number of
snapshots.} \label{sinrxsnap}
\end{center}
\end{figure}

We consider an example of adaptive beamforming in a non-stationary
scenario where the system has $6$ users with equal power and the
environment experiences a sudden change at time $i=800$. We assess
the performance of the system in terms of the
signal-to-interference-plus-noise ratio (SINR), which is defined
as
\begin{equation}
{\rm SINR} = \frac{\hat{\mathbf w}^H_D[i]\hat{\mathbf
S}_D^H[i]{\mathbf R}_s[i] \hat{\mathbf S}_D[i] \hat{\mathbf
w}_D[i]}{\hat{\mathbf w}^H_D[i]\hat{\mathbf S}_D^H[i]{\mathbf
R}_I[i] \hat{\mathbf S}_D[i] \hat{\mathbf w}_D[i]},
\end{equation}
where ${\mathbf R}_s$ denotes the covariance matrix of the signal
of interest (SoI) and ${\mathbf R}_I$ is the covariance matrix of
the interference. In the example, we consider the full-rank, the
PC \cite{jolliffe}, the MSWF \cite{honig}, the AVF \cite{qian},
the JIO \cite{delamare_jio_spl}, the JIDF \cite{delamare_jidf} and
the MVDR that assumes perfect knowledge of the covariance matrix.
The $5$ interferers impinge on the ULA at $-60^{o}$, $- 30^{o}$,
$0^{o}$, $45^{o}$, $60^{o}$ with equal powers to the SoI, which
impinges on the array at $15^o$. At the time instant $i=800$ we
have $3$ interferers with $5$ dB above the SoI's power level
entering the system with DoAs $-45^o$, $-15^o$ and $30^o$, whereas
one interferer with DoA $45^{o}$ and a power level equal to the
SoI exits the system. The results of this example are depicted in
Fig. \ref{sinrxsnap}. The curves show that the reduced-rank
algorithms have a superior performance to the full-rank algorithm.
The best performance is obtained by the JIDF scheme, which is
followed by the JIO, the AVF, the MSWF, the PC and the full-rank
algorithms.

\subsection{Audio, speech, image, and video processing:}

Echo cancellation, prediction, compression and recognition of
multimedia signals.

\subsection{Modelling for non-linear and large problems:}

Neural networks and other bio-inspired structures, Volterra
series.

\section{New frontiers and research directions:}

We will consider the relationships between reduced-rank signal
processing and algorithms for dimensionality reduction with
emerging techniques that include compressive sensing and tensor
decompositions.

Compressive sensing techniques \cite{candes}-\cite{figueiredo} can
substantially improve the performance of sensor array processing
systems including MIMO radars by taking into account and
exploiting the sparse nature of the signals encountered in these
systems. In the literature of signal processing and information
theory it has been recently shown that the use of compressive
sensing techniques \cite{candes}-\cite{figueiredo} can provide
very significant gains in performance while requiring lower
computational complexity requirements than existing techniques due
to a smarter way of processing information. The main idea behind
compressive sensing methods is to use linear projections that
extract the key information from signals and then employ a
reconstruction algorithm based on optimization techniques. In
particular, the linear projections are intended to collect the
samples that are meaningful for the rest of the procedure and
perform significant signal compression. These linear projections
are essentially dimensionality reduction procedures. Samples with
very small magnitude that cannot be discerned from noise are
typically good candidates for elimination. This is followed by a
reconstruction algorithm that aims to recreate the original signal
from the compressed version.

\section{Concluding Remarks}

In this tutorial on reduced-rank signal processing, we reviewed
design methods and algorithms for dimensionality reduction, and
discussed a number of important applications. A general framework
based on linear algebra and linear estimation was employed to
introduce the reader to the fundamentals of reduced-rank signal
processing and to describe how dimensionality reduction is
performed on an observed discrete-time signal. A unified treatment
of dimensionality reduction algorithms was presented and used to
describe some key algorithms for dimensionality reduction and
reduced-rank processing. A number of applications were considered
as well as several examples were provided to illustrate this
important area.


\begin{thebibliography}{100}

{

\bibitem{haykin}
S. Haykin, \textit{Adaptive Filter Theory}, $4^{th}$ edition,
Prentice-Hall, 2002.

% R. Bellman, \textit{Adaptive Control Processes},
%Princeton Univ. Press, 1961.

\bibitem{scharf1} L. L. Scharf, \textit{Statistical Signal
Processing: Detection, Estimation, and Time Series Analysis}, New
York: Addison-Wesley Publishing Co., 1990.

\bibitem{scharf2} L. L. Scharf, ``The SVD and reduced-rank signal
processing," \textit{Signal Processing}, 24, pp. 111-130, November
1991.

\bibitem{hotelling}
H. Hotelling, “Analysis of a Complex of Statistical Variables
into Principal Components,” Journal of Educational Psychology,
vol. 24, no. 6/7, pp. 417–441, 498–520, September/October 1933.

\bibitem{jolliffe}
I. T. Jolliffe, \textit{Principal component analysis}, New York,
Springer Verlag, 1986 (2 ed., 2002).

\bibitem{ulf}
M. O. Ulfarsson and V. Solo, ``Sparse Variable PCA Using Geodesic
Steepest Descent", \textit{IEEE Transactions on Signal
Processing}, 2001, V.56, No. 12, December 2008, pp. 5823-5832.

\bibitem{kohonen}
T. Kohonen, S. Kaski, K. Lagus, J. Salojärvi, J. Honkela, V.
Paatero and V. Saarela, ``Self organization of massive document
collection, \textit{IEEE Transactions on Neural Networks}, vol.
11, no. 3, May 2000, pp.574-585.

\bibitem{delamaresp}
R. C. de Lamare and R. Sampaio-Neto, ``Adaptive Reduced-Rank MMSE Filtering
with Interpolated FIR Filters and Adaptive Interpolators", \textit{IEEE Sig.
Proc. Letters}, vol. 12, no. 3, March, 2005.

\bibitem{lafon}
S. Lafon and A.B. Lee, ``Diffusion maps and coarse-graining: A
unified framework for dimensionality reduction, graph
partitioning, and data set parameterization", \textit{ IEEE
Transactions on Pattern Analysis and Machine Intelligence}, vol.
28, no. 9, 2006, pp.1393–1403.

\bibitem{law}
M.H.C. Law, A.K. Jain, ``Incremental nonlinear dimensionality
reduction by manifold learning", \textit{IEEE Transactions on
Pattern Analysis and Machine Intelligence}, vol. 28, no. 3, 2006,
pp. 377 - 391.

\bibitem{yan}
Jun Yan, Benyu Zhang, Ning Liu, Shuicheng Yan, Qiansheng Cheng, W.
Fan, Qiang Yang, W. Xi, Zheng Chen, ``Effective and efficient
dimensionality reduction for large-scale and streaming data
preprocessing ", \textit{IEEE Transactions on Knowledge and Data
Engineering}, vol. 18, no. 3, 2006 , pp. 320 - 333.

\bibitem{sanguinetti}
G. Sanguinetti, ``Dimensionality Reduction of Clustered Data
Sets", \textit{IEEE Transactions on Pattern Analysis and Machine
Intelligence}, vol. 30, no. 3, 2008, pp. 535-540.

\bibitem{cga} M. R.
Hestenes and E. Stiefel, “Methods of Conjugate Gradients for
Solving Linear Systems,” Journal of Research of the National
Bureau of Standards, vol. 49, no. 6, pp. 409–436, December 1952.

\bibitem{lanczos}
C. Lanczos, “Solution of Systems of Linear Equations byMinimized
Iterations,” Journal of Research of the National Bureau of
Standards, vol. 49, no. 1, pp. 33– 53, July 1952.

\bibitem{arnoldi}
W. E. Arnoldi, “The Principle of Minimized Iterations in the
Solution of the Matrix Eigenvalue Problem,” Quarterly of Applied
Mathematics, vol. 9, no. 1, pp. 17–29, January 1951.

\bibitem{csmetric}
J. S. Goldstein and I. S. Reed, “Reduced-Rank Adaptive
Filtering,” IEEE Transactions on Signal Processing, vol. 45, no.
2, pp. 492–496, February 1997.


\bibitem{goldstein} J. S. Goldstein, I. S. Reed, and L. L. Scharf,
``A multistage representation of the Wiener filter based on
orthogonal projections," \textit{IEEE Transactions on Information
Theory}, vol. 44, November 1998.

\bibitem{Badeau}
R. Badeau, B. David, and G. Richard, ``Fast approximated power
iteration subspace tracking," \textit{IEEE Trans. Signal
Processing}, vol. 53, pp. 2931-2941, Aug. 2005.

\bibitem{honig}
M. L. Honig and J. S. Goldstein, ``Adaptive reduced-rank
interference suppression based on the multistage Wiener filter,"
\textit{IEEE Transactions on Communications}, vol. 50, no. 6, June
2002.

\bibitem{delamare_ccmmswf}
R. C. de Lamare, M. Haardt, and R. Sampaio-Neto, ``Blind Adaptive
Constrained Reduced-Rank Parameter Estimation based on Constant
Modulus Design for CDMA Interference Suppression", \textit{IEEE
Transactions on Signal Processing}, June March 2008.

\bibitem{song}
N. Song, R. C. de Lamare, M. Haardt, and M. Wolf, ``Adaptive Widely Linear
Reduced-Rank Interference Suppression based on the Multi-Stage Wiener Filter,"
IEEE Transactions on Signal Processing, vol. 60, no. 8, 2012.

\bibitem{pados}
D. A. Pados and G. N. Karystinos, ``An iterative algorithm for the
computation of the MVDR filter," \textit{IEEE Trans. on Sig.
Proc.}, vol. 49, No. 2, February, 2001.

\bibitem{qian}
H. Qian and S.N. Batalama, ``Data record-based criteria for the
selection of an auxiliary vector estimator of the MMSE/MVDR
filter", \textit{IEEE Trans. on Communications}, vol. 51, no. 10,
Oct. 2003, pp. 1700 - 1708.

\bibitem{hua1} Y. Hua and M. Nikpour,
``Computing the reduced rank Wiener filter by IQMD," \textit{IEEE
Signal Processing Letters}, pp. 240-242, Vol. 6, Sept. 1999.

\bibitem{hua} Y. Hua, M. Nikpour, and P. Stoica, ``Optimal
reduced-rank estimation and filtering," \textit{IEEE Transactions
on Signal Processing}, 2001, V.49,
457-469.

\bibitem{delamare_jio_spl} R. C. de Lamare and R.
Sampaio-Neto, ``Reduced-Rank Adaptive Filtering Based on Joint
Iterative Optimization of Adaptive Filters, " \textit{IEEE Signal
Processing Letters}, Vol. 14 No. 12, December 2007, pp. 980 - 983.

\bibitem{delamaretvt10}
R. C. de Lamare and R. Sampaio-Neto, ``Reduced-Rank Space-Time Adaptive
Interference Suppression With Joint Iterative Least Squares Algorithms for
Spread-Spectrum Systems," \textit{IEEE Transactions on Vehicular Technology},
vol.59, no.3, March 2010, pp.1217-1228.

\bibitem{jio_mimo}
R.C. de Lamare and R. Sampaio-Neto, “Adaptive reduced-rank equalization
algorithms based on alternating optimization design techniques for MIMO
systems,” IEEE Trans. Veh. Technol., vol. 60, no. 6, pp. 2482-2494, July 2011.

\bibitem{delamare_jidf}
R. C. de Lamare and R. Sampaio-Neto, ``Adaptive Reduced-Rank
Processing Based on Joint and Iterative Interpolation, Decimation
and Filtering", \textit{IEEE Transactions on Signal Processing},
vol. 57, no. 7, July 2009, pp. 2503 - 2514.

\bibitem{fa10}
R. Fa, R. C. de Lamare, and L. Wang, ``Reduced-Rank STAP Schemes for Airborne
Radar Based on Switched Joint Interpolation, Decimation and Filtering
Algorithm," \textit{IEEE Transactions on Signal Processing}, vol.58, no.8, Aug.
2010, pp.4182-4194.

\bibitem{barc}
R.C. de Lamare, R. Sampaio-Neto, M. Haardt, "Blind Adaptive Constrained
Constant-Modulus Reduced-Rank Interference Suppression Algorithms Based on
Interpolation and Switched Decimation," \textit{IEEE Transactions on Signal
Processing}, vol.59, no.2, pp.681-695, Feb. 2011.

\bibitem{aic}
H. Akaike, ``A new look at the statistical model identification,"
IEEE Transactions on Automatic Control, vol. 19, no. 6, pp.
716–723, 1974.

\bibitem{mdl}
J. Rissanen, ``Modeling by shortest data description,
\textit{Automatica}, vol. 14, pp. 465-471, 1978.

\bibitem{stoicaspm}
P. Stoica and Y. Selén, ``Model-order selection: a review of
information criterion rules," IEEE Signal Processing Magazine,
vol. 21, no. 4, pp. 36–47, 2004.

\bibitem{xiao}
W. Xiao and M. L. Honig, ``Large System Transient Behavior of
Adaptive Least Squares Algorithms'', IEEE Transactions on
Information Theory, Vol. 51, No. 7, pp. 2447-2474, July 2005.

\bibitem{sun}
Z. Sun and S. S. Ge, \textit{Switched Linear Systems: Control and
Design}, London: Springer-Verlag, 2005.

\bibitem{schizas}
I.D. Schizas, G.B. Giannakis, Z.-Q. Luo, ``Distributed Estimation
Using Reduced-Dimensionality Sensor Observations", IEEE
Transactions on Signal Processing, vol. 55, no. 8, 2007, pp.
4284-4299.

%\bibitem{bert}
%D. P. Bertsekas, {\it Nonlinear Programming}, Athena Scientific,
%2nd Ed., 1999.


\bibitem{luen}
D. Luenberger, \textit{Linear and Nonlinear Programming}, 2nd Ed.
Addison-Wesley, Inc., Reading, Massachusetts 1984.

\bibitem{candes}
E. Candès and M. B. Wakin, "An Introduction to Compressive
Sampling", IEEE Signal Processing Magazine, March 2008, pp. 21 -
30.

\bibitem{donoho}
D. Donoho, "Compressed Sensing," IEEE Trans. Information
Theory, vol. 52, no. 4, pp. 1289-1306.

\bibitem{candes}
E. Candès and T. Tao, "Near-optimal Signal Recovery from
Random Projections: Universal Encoding Strategies?, "IEEE Trans.
Information Theory, vol. 52, no. 12, pp. 5406-5425.

\bibitem{haupt}
J. Haupt and R. Nowak, "Signal Reconstruction from Noisy
Random Projections", IEEE Trans. Information Theory, vol. 52, no.
9, pp. 4036-4048.

\bibitem{figueiredo}
M. Figueiredo, R. Nowak, S. Wright, "Gradient projection
for sparse reconstruction: application to compressed sensing and
other inverse problems",  IEEE Journal of Selected Topics in
Signal Processing: Special Issue on Convex Optimization Methods
for Signal Processing, vol. 1, no. 4, pp. 586-598, 2007.


}
\end{thebibliography}
\end{document}